\documentclass[11pt,a4paper]{article}
\usepackage[inline]{enumitem}
\usepackage{amsfonts}
\usepackage{amsmath,amssymb,amsthm}
\usepackage{mathtools}
\usepackage{graphicx,subfigure}
\usepackage[margin=1in,a4paper]{geometry}
\usepackage{algorithm}
\usepackage{algorithmic}
\usepackage{xcolor}
\usepackage{natbib}
\usepackage{tabu}
\usepackage[colorlinks,citecolor=blue,urlcolor=blue,bookmarks=false,hypertexnames=true]{hyperref} 
\usepackage{newunicodechar}

\usepackage{mathtools}
\usepackage{setspace}
\mathtoolsset{showonlyrefs=true}
\newtheorem{proposition}{Proposition}

\theoremstyle{remark}
\newtheorem{remark}{Remark}

\allowdisplaybreaks[4]

\title{A Two-Step Framework for Arbitrage-Free Prediction of the Implied Volatility Surface} 

\author{Wenyong Zhang\thanks{Department of Systems Engineering and Engineering Management, The Chinese University of Hong Kong, Hong Kong SAR. Email: wyzhang@se.cuhk.edu.hk.}, Lingfei Li\thanks{Corresponding author. Department of Systems Engineering and Engineering Management, The Chinese University of Hong Kong, Hong Kong SAR. Email: lfli@se.cuhk.edu.hk.}, Gongqiu Zhang\thanks{School of Science and Engineering, The Chinese University of Hong Kong, Shenzhen, China. Email: zhanggongqiu@cuhk.edu.cn.}}

\date{December 26, 2021}

\begin{document}

\maketitle

\begin{abstract}
  We propose a two-step framework for predicting the implied volatility surface over time without static arbitrage. In the first step, we select features to represent the surface and predict them over time. In the second step, we use the predicted features to construct the implied volatility surface using a deep neural network (DNN) model by incorporating constraints that prevent static arbitrage. We consider three methods to extract features from the implied volatility data: principal component analysis, variational autoencoder and sampling the surface, and we predict these features using LSTM. Using a long time series of implied volatility data for S\&P500 index options to train our models, we find two feature construction methods, sampling the surface and variational autoencoders combined with DNN for surface construction, are the best performers in out-of-sample prediction. In particular, they outperform a classical method substantially. Furthermore, the DNN model for surface construction not only removes static arbitrage, but also significantly reduces the prediction error compared with a standard interpolation method. Our framework can also be used to simulate the dynamics of the implied volatility surface without static arbitrage. 
  
  \bigskip
\end{abstract}

\section{Introduction}\label{sec:intro}
The implied volatility for an option with given strike $K$ and time to maturity $\tau$ is the volatility level that makes the option price from the Black-Scholes formula equal to the observed option price. In practice, traders quote implied volatility for the price of call and put options. The implied volatility surface (IVS) is the collection of implied volatilities as a function of $K$ and $\tau$, and it is a fundamental input for various tasks, such as derivatives pricing and hedging, volatility trading, and risk management. 

There exists many studies on the implied volatility surface. Since only a finite set of options are traded on each day, an important task in practice is to interpolate and extrapolate the implied volatitlies for these options to obtain the entire surface. Methods for this task mainly fall into three categories: 

(a) Parametric models: \cite{gatheral2004parsimonious} proposes the well-known stochastic volatility inspired (SVI) model for single-maturity implied volatility skews and \cite{gatheral2014arbitrage} further generalizes the model to obtain the surface without static arbitrage. 

(b) Splines: \cite{fengler2009arbitrage}, \cite{orosi2015arbitrage} and \cite{fengler2015semi} develop spline-based models considering the no static arbitrage conditions. 

(c) Machine learning models: Neural networks have been utilized to construct the IVS. See \cite{zheng2019gated} for a neural network model that combines several single layer neural networks and \cite{deep_IVS} for a neural network model with multiple layers and hence deep. Both works formulate the static no arbitrage constraints as penalties in their loss functions to obtain a surface free of static arbitrage. \cite{bergeron2021variational} employs the variational autoencoder for constructing the IVS. This method first extracts latent factors for observed implied volatilities through a neural network called encoder and then obtains the surface through another neural network known as decoder by using the latent factors together with moneyness and time to maturity as inputs. \cite{almeida2021can} boosts the performance of classical parametric option pricing models by fitting a neural network to the residuals of these parametric models. \cite{horvath2021deep} uses a deep neural network to approximate the pricing function of a sophisticated rough stochastic volatility model to facilitate its calibration to implied volatility data.

In this paper, we focus on the dynamic prediction problem of IVS, i.e., at time $t$ we predict the IVS of time $t'$ with $t'>t$. While interpolation and extrapolation of the IVS for a single day has been well studied with satisfactory solutions, the dynamic prediction problem requires more research in our opinion. It can also be more challenging than the former problem as we need to capture the behavior of IVS over time. Unlike predicting financial quantities like stock prices, interest rates and exchange rates, predicting the IVS is a more challenging problem as we need the whole surface which is a bivariate function as opposed to e.g., the price of a stock which is a single value. Furthermore, the predicted surface must satisfy certain restrictions so that it is free of arbitrage, whereas there is no such concern for predicting fundamental financial variables.   

\subsection{Some Related Literature}

The dynamics of the IVS has been analyzed in various papers. \cite{skiadopoulos2000dynamics}, \cite{fengler2003dynamics} and \cite{cont2002dynamics} apply principal component analysis (PCA) to IVS data. The first two papers perform PCA for smiles of different maturities while the third reference performs PCA directly on the surface. It is found in \cite{cont2002dynamics} that the first three eigenmodes already explain most of the variations and they are associated with the level, skewness and convexity of the IVS. The authors approximate the IVS on each day using the linear combination of the first three eigensurfaces and the dynamics of each coefficient is modelled by a first-order autoregressive model. \cite{fengler2007semiparametric} develop a semiparametric factor model for the IVS, where they assume the IVS is given by a sum of basis functions, each of which uses moneyness and time to maturity as inputs. The factor loadings are modelled by a vector autoregressive process. The basis functions and the loadings are estimated by weighted least squares with the weight given by a kernel function. 
\cite{bloch2019neural} proposes a general modeling framework. He uses several risk factors to represent the surface and model each parameter (corresponding to a risk factor) using a neural network. In particular, he provides three ways of obtaining the risk factors: using the first three eigenmodes from PCA in \cite{cont2002dynamics} or the parameters of a polynomial model for the IVS, or the parameters of a stochastic volatility type model (such as the SVI model). Explanatory variables such as the spot price, volume, VIX, etc can be input into the neural networks for these parameters. \cite{cao2020neural} model the relationship between the expected daily change in the implied volatility of S\&P500 index options by a multilayer feedforward neural network with daily index return, VIX, moneyness and time to maturity as inputs. They find that with the index return and VIX as features, their model can improve an analytic model significantly. One can use all of these models for predicting the IVS on a future date, but their predicative performance is not assessed in these papers. Furthermore, all these papers do not show how to obtain the dynamics of the IVS without arbitrage. 

Several other papers directly address the dynamic prediction problem. \cite{dellaportas2014arbitrage} predict the implied volatilities of a single maturity by forecasting the parameters in the SABR model from a time series model. As the SABR model is arbitrage free, the predicted implied volatility smile has no arbitrage. However, their approach does not apply to the whole surface because the SABR model cannot fit the surface well. 
\cite{goncalves2006predictable} and \cite{bernales2014can} model the IVS as a polynomial of time-adjusted moneyness and time to maturity. To predict the IVS on a future date, they use the forecasted coefficients of the polynomial from a time series model. 
\cite{audrino2010semi} develop a regression tree model through boosting to predict the IVS.
\cite{chen2019forecasting} apply the long short-term memory (LSTM) model with attention mechanism to predict the IVS. \cite{bloch2020predicting} predicts the IVS by predicting the risk factors driving the dynamics of the IVS using temporal difference backpropagation (TDBP) models.  \cite{zeng2019online} use tick data on options to construct the implied volatilty surface at high frequency through a support vector regression model. A common issue with all these papers is that the predicted IVS from their models is not necessarily arbitrage free. Recently, \cite{ning2021arbitrage} introduce an interesting approach to simulate arbitrage-free IVS over time. They first calibrate an arbitrage-free stochastic model to the IVS data and then generate the model parameters of a future date from a variational autoencoder. The future IVS is then obtained under the stochastic model using the generated model parameters. However, they don't use their approach to predict future IVS.

\subsection{Our Contributions}
The contributions of this paper are twofold. First, we provide a general framework for dynamic prediction and simulation of IVS free of static arbitrage. This is a new feature provided by our approach compared with existing methods for dynamic prediction of IVS.
Second, we show how to construct features to represent the IVS and develop some successful models for predicting the IVS in this framework. 

Our framework consists of two steps. 
\begin{itemize}
	\item Step 1: We select features to represent the IVS and predict them. The predicted features are mapped to a discrete set of implied volatilities. If the task is simulation, we simulate the features in this step. 
	
	\item Step 2: We construct the entire IVS from the discrete set of implied volatilities in Step 1 through a deep neural network (DNN) model by taking into account the constraints that rule out static arbitrage. 
\end{itemize}

This framework is completely flexible as users can construct features and predict them in their own ways. Furthemore, any results obtained in Step 1 can be converted to a static-arbitrage free surface through the DNN model in Step 2. However, the predicted surface from our framework may admit dynamic arbitrage opportunities as we don't enforce constraints that prevent dynamic arbitrage in our model. It would be difficult to do so in our framework as it is data-driven and assumes no stochastic model for the underlying asset.

The accuracy of predicting the IVS obviously hinges on the selected features and the model for prediciting them. In general, one can use features extracted from the IVS data together with exogenous variables to represent the surface. The focus of this paper is on how to extract features and we consider three approaches: using the eigenmodes from the PCA analysis of \cite{cont2002dynamics}, applying the variational autoencoder (VAE) to extract latent factors for the IVS, and directly sampling the IVS on a discrete grid of moneyness and time to maturity. PCA can be viewed as a parametric approach that approximates the change of the log-IVS using a linear combination of eigensurfaces.
The VAE approach is more general than the PCA as it offers a flexible nonlinear factor representation of the surface. The sampling approach is completely nonparametric. 

To predict the extracted features, we utilize the long short-term memory (LSTM) model, which is a popular deep learning model for sequential data. Our choice is motivated by the success of deep learning in a range of prediction problems in finance. See e.g., \cite{borovykh2017conditional} and \cite{sezer2020financial} for various financial time series, \cite{sirignano2019deep} and \cite{sirignano2019universal} for limit order books, \cite{sirignano2018deep} for mortgage risk and \cite{yan2018} for tail risk in asset returns. 

By training our models using data of 9.5 years and putting them to test in a period of 2.5 years, we find that both the sampling and the VAE approach are quite successful in predicting the IVS. The error of the PCA approach is almost three times of the other two, indicating that prediction based on a linear combination of eigensurfaces is not accurate enough.

Another important finding is that the DNN model in Step 2 not only serves the purpose of constructing an arbitrage free surface but it is also crucial for prediction accuracy. Compared with a standard interpolation method for the IVS, using the DNN model can reduce the out-of-sample prediction error substantially for the sampling and VAE approach. 

Our paper is related to \cite{bloch2019neural} which also provides a general and appealing framework, but they differ in a number of ways. First, \cite{bloch2019neural} doesn't consider how to obtain arbitrage free surfaces. Second, \cite{bloch2019neural} constructs features for the IVS using PCA and parametric models (such as the polynomial or SVI model). We offer another two approaches for feature construction in this paper. Our empirical results suggest that the prediction model based on features from PCA is not good enough and we also have some potential issues with predicting features from parametric models (see Remark \ref{remk:parametric-models}). Third, \cite{bloch2019neural} proposes to model each feature by a neural network. We model these features jointly by one model (LSTM in our implementation). Having separate neural network models for the features may fail to capture potential dependence among them unless they are independent. Finally, \cite{bloch2019neural} doesn't provide empirical study to validate his models.

The rest of the paper is organized as follows. Section \ref{sec:IVS} provides background information on the implied volatility surface, including the definition of implied volatility, conditions that ensure no static arbitrage, an interpolation method and our data. Section \ref{sec:framework} presents the two-step framework for prediction and simulation. Section \ref{sec:empirical} shows empirical results and compares different models. Section \ref{sec:conclusion} concludes with remarks for future research.

\section{The Implied Volatility Surface}\label{sec:IVS}

We provide some background knowledge on the implied volatility surface (IVS) in this section. 

\subsection{Implied Volatility}
Consider a European call option on a dividend paying asset $S_{t}$ with maturity date $T$ and strike price $K$. Set $\tau = T-t$, which is the time to maturity. Denote the risk-free rate by $r$ and the dividend yield by $d$. Let $F_t(\tau)$ be the forward price at $t$ for time to maturity $\tau$. It is given by $F_{t}(\tau) = S_{t}\mathrm{e}^{(r-q)\tau}$. We will write $F_t$ below to simplify the notation. 

Under the Black-Scholes model, the call option price at time $t$ is given by
\begin{equation}
	C\left(F_{t}, K, \tau, \sigma\right)=\mathrm{e}^{-r \tau}\left(F_{t} N\left(d_{1}\right)-K N\left(d_{2}\right)\right),
\end{equation}
where 
\begin{align} 
	d_{1}&=\frac{-m+\frac{1}{2} \sigma^{2} \tau}{\sigma\sqrt{\tau}},\ d_{2}=\frac{-m-\frac{1}{2} \sigma^{2} \tau}{\sigma\sqrt{\tau}},  \label{d}\\
    m &= \ln \left(\frac{K}{F_{t}}\right),  \label{m}
\end{align}
and $N(\cdot)$ is the cumulative distribution function of the standard normal distribution. The variable $m$ defined in \eqref{m} is known as the log forward moneyness.

It is well-known that the Black-Scholes model is unrealistic. To use it in practice, one looks for the level of volatility to match an observed option price, i.e., we solve $\sigma$ from the equation
\begin{equation}
	C\left(F_{t}, K, \tau, \sigma\right)=C_{mkt}, 
\end{equation}
where $C_{mkt}$ is the observed market price for the call option. The solution is called the implied volatility. 

The implied volatility surface at a time point is the collection of implied volatilities for options with different $K$ and $\tau$.  We prefer to consider the IVS as a function of $m$ and $\tau$, because $m$ is a relative coordinate. Hereafter, the IVS at time $t$ is denoted by $\sigma_t(m,\tau)$. Practitioners also like to quote implied volatilities using the Black-Scholes Delta (detnoted by $\delta$) and $\tau$, where 
$\delta=e^{-q\tau}N(d_1)$.

\subsection{Static Arbitrage Free Conditions}
Conditions that ensure the implied volatility surface is free of static arbitrage have been well studied in the literature (see e.g., \cite{roper2010arbitrage}, \cite{gulisashvili2012analytically}). We summarize them in the proposition below. 

\begin{proposition}\label{prop: NA}
Consider an IVS $\sigma(m,\tau)$ and suppose the following conditions are satisfied:
\begin{itemize} 
\item[1.] \textbf {(Positivity)} $\sigma(m, \tau)>0$ for every $(m, \tau)$.

\item[2.] \textbf{(Twice Differentiability)} For every $\tau>0, m \rightarrow \sigma(m, \tau)$ is twice differentiable.

\item[3.] \textbf{(Monotonicity)} For every $m \in \mathbb{R}, \tau \rightarrow \sigma(m, \tau)^{2}\tau$ is increasing, or equivalently
\begin{equation}\label{cal-NA}
\ell_{\mathrm{cal}}(m, \tau) = \sigma(m, \tau)+2 \tau \partial_{\tau} \sigma(m, \tau) \geq 0.
\end{equation}

\item[4.] \textbf{(Durrleman’s Condition)} For every $(m, \tau)$,
\begin{equation}\label{but-NA}
	\ell_{\mathrm{but}}(m, \tau) = \left(1-\frac{m \partial_{m} \sigma(m, \tau)}{\sigma(m, \tau)}\right)^{2}- \frac{\left(\sigma(m, \tau) \tau \partial_{m} \sigma(m, \tau)\right)^{2}}{4} +\tau \sigma(m, \tau) \partial_{m m} \sigma(m, \tau) \geq 0. 
\end{equation}

\item[5.] \textbf{(Large Moneyness Behavior)} For every $\tau$, $\sigma^2(m, \tau)$ is linear as $|m| \rightarrow + \infty$.
\end{itemize}
Then $\sigma(m,\tau)$ is free of static arbitrage. 
\end{proposition}

Condition 3 implies that $\sigma(m, \tau)$ is free of calendar spread arbitrage and condition 4 guarantees that there is no butterfly arbitrage. In Section \ref{sec:loss}, we will implement these conditions to get an arbitrage free surface. 

\subsection{Interpolation for the Implied Volatility Surface}
On a given day, implied volatilities can only be calculated for a discrete set of $(m,\tau)$ pairs, which correspond to options that are listed on that day. Suppose we are given $\{\sigma(m_i,\tau_i): i=1,\cdots,n\}$ on a day. There are various approaches to interplate these given points to obtain the entire IVS. Here, we consider a simple and popular parametric model proposed in \cite{dumas1998implied} and hereafter it will simply be called DFW. This model assumes
\begin{equation}\label{poly}
	\sigma(m, \tau) = \max(0.01, a_0 + a_1 m + a_2 \tau + a_3 m^2 + a_4 \tau^2 + a_5 m\tau),
\end{equation}
where a floor of $0.01$ is set to prevent the implied volatility of the model from becoming too small or even negative. The coefficients $a_0,\cdots,a_5$ can be estimated by regression. 

Another popular non-parametric approach to estimate the entire implied volatility surface uses the Nadaraya–Watson (NW) estimator (\cite{hardle1990applied}), which is given by
\begin{equation} \label{nw}
\bar{\sigma}_{t}(m, \tau)=\frac{\sum_{i=1}^{n} \sigma_{t}\left(m_{i}, \tau_{i}\right) g\left(m-m_{i}, \tau-\tau_{i}\right)}{\sum_{i=1}^{n} g\left(m-m_{i}, \tau-\tau_{i}\right)},
\end{equation}
where 
\begin{equation}\label{gauss-kernel}
g(x, y)=\frac{1}{2\pi}\exp \left(-\frac{x^{2}}{2 h_{1}}\right) \exp \left(-\frac{y^{2}}{2 h_{2}}\right)
\end{equation} 
is a Gaussian kernel. The estimator involves two hyper-parameters $h_{1}$ and $h_{2}$, which are bandwidths and they determine the degree of smoothing. If the parameters are too small, a bumpy surface is generated. If they are too big, important details in the observed data can be lost after smoothing. When implementating this approach, on each day we apply five-fold cross-validation to the implied volatility data of this day to select the best pair of $(h_1,h_2)$ from a grid of values for them. 

Table \ref{table_nwdfw} presents the average root mean squared error (RMSE) of interpolation of these two methods, where the average is taken over days in our training period. The DFW model is more accurate and it will be our choice for further study. The larger error of the NW estimator is very likely caused by applying the same bandwidth $(h_1,h_2)$ to all points, whereas our implied volatility data is non-uniformly distributed in the $(m,\tau)$ space.
\begin{table}[htbp!]
	\centering
	\begin{tabular}{ccccc}
		\hline
		& DFW       & NW \\
		\hline
		RMSE       & 0.018 & 0.026 \\		
		\hline
	\end{tabular}%
	\caption{The average RMSE for NW and DFW}
	\label{table_nwdfw}%
\end{table}%

\subsection{Data}

\begin{figure}[htbp!]
	\centering
	\subfigure[the surface]{
	\begin{minipage}[t]{0.48\textwidth}
		\centering
		\includegraphics[scale=0.45]{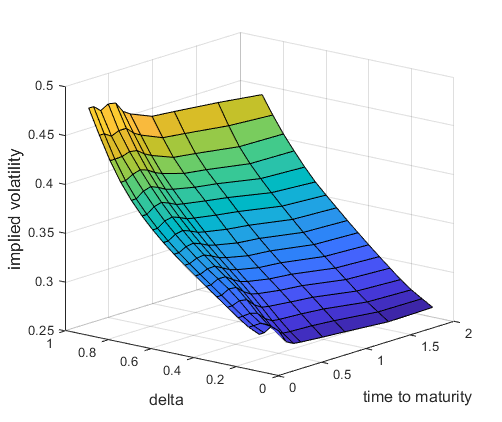}
	\end{minipage}
	}
	\subfigure[the skews for various maturities]{
	\begin{minipage}[t]{0.48\textwidth}
		\centering
		\includegraphics[scale=0.35]{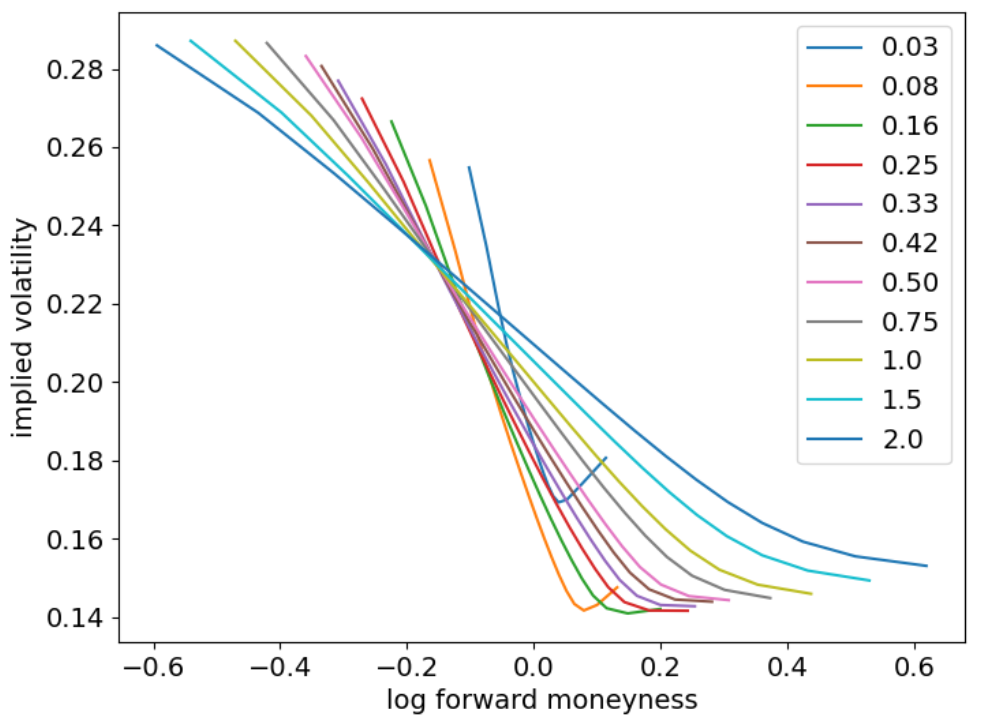}
	\end{minipage}
	}
	\caption{The average implied volatility surface of S\&P500 index options with maturity up to two years and the skews of different maturities. In plot (b), we show the skews as functions of log forward moneyness. }\label{fig:mean-IVS}
\end{figure}

The dataset used for this paper contains implied volatility surfaces for the S\&P500 index options from January 1, 2009 to December 31, 2020. We obtained the data from OptionMetrics through 
the Wharton Research Data Services. On each day, we have implied volatilities for a set of $(\delta,\tau)$ pairs with $\delta$ going from $0.1$ to $0.9$ with an increment of $0.05$ and $\tau=10, 30, 60, 91, 122, 152, 182, 273, 365, 547, 730$ calendar days. Since we consider the IVS as a function of $m$ and $\tau$, we convert $\delta$ to $m$ using
\begin{equation}
	m = \frac{1}{2}\sigma^2 \tau -  \sigma\sqrt{\tau}N^{-1}\left(e^{q\tau}\delta\right).
\end{equation}
This results in implied volatilies on different days for different grids of moneyness but the same grid of $\tau$. We denote by $\mathcal{I}_{d,t}$ the set of $(m,\tau)$ pairs for observed implied volatilities at time $t$. In total, we have data on 3021 days and on each day 374 points are observed from the implied volatility surface (i.e., the size of $\mathcal{I}_{d,t}$ is 374). To demonstrate salient features of the implied volatility surface for index options, we calculate the average of $\sigma_t(\delta,\tau)$ over $t$ for all observed $(\delta,\tau)$ pairs, and plot the average values as a surface in Figure \ref{fig:mean-IVS}(a) in terms of $\delta$ and $\tau$. In Figure \ref{fig:mean-IVS}(b), we show the implied volatility curves for different maturities as functions of log forward moneyness.  A volatility skew is clearly observed for each $\tau$ and it remains quite steep even for large maturities.

As the methods we will apply later cannot be used if the $(m,\tau)$-grid changes from day to day, we have to fix it. We use the following grid for $(m,\tau)$:
\begin{equation}\label{sample-grid}
	\mathcal{I}_{0}=\left\{(m, \tau): m \in \mathcal{M}_{0}, \tau \in \mathcal{T}_{0}\right\},
\end{equation}
where
\begin{align}
\mathcal{M}_{0}&=\{\log (x): x \in\{0.6, 0.8, 0.9, 0.95, 0.975, 1, 1.025, 1.05, 1.1, 1.2, 1.3, 1.5, 1.75, 2\}\}, \\
\mathcal{T}_{0}&=\{i / 365: i \in\{10, 30, 60, 91, 122, 152, 182, 273, 365, 547, 730\}\}.
\end{align}
Since the set of $\tau$ is fixed over time in the data, we simply adopt the set as the grid for $\tau$ but change the time unit to year ($\tau$ was quoted in days initially). For the grid of moneyness, we first get the minimum value and maximum value of $m$ in our dataset and set $[\log(0.6),\log(2)]$ as the range, which is slightly wider than the range from the minimum to the maximum. Then we create a non-uniform grid on this range so that the grid is denser near ATM. As $\mathcal{I}_{0}$ is different from the observed grid for $(m,\tau)$ on a day, we use the DFW model given in \eqref{poly} to obtain the implied volatilities on $\mathcal{I}_{0}$. Eventually, at every $t$, we have a set of 154 implied volatilities
\begin{equation}\label{sample-IVS}
	\bar{\Sigma}_t=\{\bar{\sigma}_t(m,\tau): (m,\tau)\in \mathcal{I}_{0}\},
\end{equation}
which can be deemed as a sample of the implied volatility surface.

\section{A Two-Step Framework}\label{sec:framework}
Consider the implied volatility surface process $\{\sigma_t(m,\tau), t\ge 0\}$. We would like to predict $\sigma_{T+1}(m,\tau)$ (the entire surface) given the information available at $T$. In reality, we do not observe the entire IVS on a day, but only the implied volatilities for a finite number of $(m,\tau)$ pairs. Furthermore, the observed $(m,\tau)$ pairs can vary from day to day. Another important problem is how we can ensure the predicted surface is free of static arbitrage. We propose a two-step framework to deal with these problems. 

\begin{figure}[htbp!]
	\centering
	\includegraphics[scale=0.6]{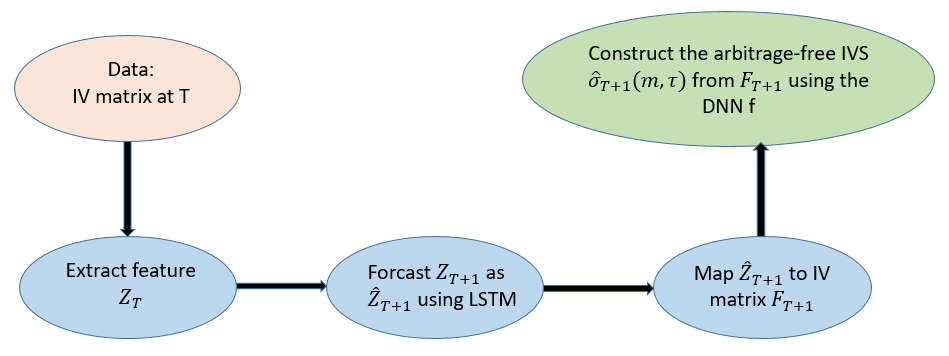}
	\caption{The workflow of the two-step framework}
	\label{fig:workflow}
\end{figure}

\begin{itemize}
	\item Step 1: We select a feature vector $Z_t$ to represent $\sigma_t(m,\tau)$ for every $t$. Given $\{Z_0,\cdots,Z_T\}$, we predict $Z_{T+1}$ using some model and the predictor is denoted by $\hat{Z}_{T+1}$. 
	
	\item Step 2:  We map $\hat{Z}_{T+1}$ to $F_{T+1}$, a discrete set of implied volatilities for $T+1$, using some function $h$, i.e, $F_{T+1}=h(\hat{Z}_{T+1})$. We predict the implied volatility surface at $T+1$ as $\hat{\sigma}_{T+1}(m,\tau)=f(m,\tau,F_{T+1})$, where $f$ is a deep neural network (DNN) that outputs an implied volatility surface free of static arbitrage.     
\end{itemize}
A flowchart is provided in Figure \ref{fig:workflow} to illustrate the framework.

In this paper, we focus on the day-ahead prediction but our framework can obviously be applied to predict the IVS for any time horizon by replacing $T+1$ with $T+m$ where $m$ is the length of the prediction horizon. The framework is flexible enough to accommodate various features and different ways of predicting them. We will explore some choices in this paper. The function $h$ can be determined according to the selected features. 

\subsection{Feature Extraction}\label{sec:feature}
We consider several methods to extract features from the implied volatility data. 

\smallskip
\noindent \textbf{Method 1 (SAM)}: We directly use the sampled implied volatility set $\bar{\Sigma}_t$ (see \eqref{sample-IVS}) to represent the entire implied volatility surface. Thus, the feature vector $Z_t=\bar{\Sigma}_t$, which is a $154$ dimensional vector in our data. We name this method as the sampling approach or SAM for short. Here the function $h$ is the identity fuction, i.e., $F_{T+1}=\hat{Z}_{T+1}$, because the predicted $\hat{Z}_{T+1}$ is a set of implied volatilities. 

\smallskip
While having a high-dimensional feature vector can better approximate the surface, predicting it may be more difficult. Thus, natually we can consider some dimension reduction techniques to extract features, which leads us to the following two methods. 

\smallskip
\noindent \textbf{Method 2 (PCA)}: \cite{cont2002dynamics} applied the surface principle component analysis (PCA) to dissect the dynamics of implied volatility surfaces. We follow their approach here. As a fixed $(m,\tau)$-grid is needed, we consider $\{\bar{\Sigma}_t, t\ge 0\}$. Define 
$X_{t}(m, \tau)=\ln \bar{\sigma}_{t}(m, \tau)$, where $\bar{\sigma}_{t}(m, \tau)\in \bar{\Sigma}_t$ and 
\begin{equation}
	U_{t}(m, \tau)=\ln \bar{\sigma}_{t}(m, \tau)-\ln \bar{\sigma}_{t-1}(m, \tau)\quad \text{for}\ (m,\tau)\in\mathcal{I}_0.
\end{equation}
Then we perform principle component analysis on $\{U_{t}(m, \tau), (m,\tau)\in\mathcal{I}_0\}$, which is a 154 dimensional random vector in our data. Let $\textbf{K}$ be the covariance matrix with $K\left(x_{1}, x_{2}\right)=\operatorname{cov}\left(U\left(x_{1}\right), U\left(x_{2}\right)\right)$, $x_{1}, x_{2} \in \mathcal{I}_{0}$. We solve the eigenvalue problem
\begin{equation}
	\textbf{K}f_{k} = v_{k} f_{k}
\end{equation}
where $v_k\ge 0$ is the $k$-th eigenvalue and $f_k$ is the associated normalized eigenvector. We sort the eigenvalues in a descending order and use the linear combination of the first $K$ eigenvectors to approximate $X_t(m, \tau)-X_0(m, \tau)$, which is 
\begin{equation}\label{pca-x}
	X_t(m,\tau)-X_0(m, \tau)\approx \sum_{k=1}^K x_k(t) f_k(m,\tau),
\end{equation}
where the coefficient
\begin{equation}
	x_k(t)=\left\langle X_{t}-X_0, f_{k}\right\rangle,
\end{equation}
the inner product between the vectors $X_{t}-X_0$ and $f_{k}$. Consequently, we have
\begin{equation}
\bar{\sigma}_{t}(m, \tau)\approx\bar{\sigma}_{0}(m, \tau) \exp \left(\sum_{k=1}^K x_{k}(t) f_{k}(m, \tau)\right).
\end{equation}
Thus, we have $Z_t=(x_1(t),\cdots,x_K(t))$ as the feature vector for $\sigma_t(m,\tau)$. Typically, a small $K$ already explains most of the variation in the data, so the feature vector is low dimensional. Let $\hat{x}_k(T+1)$ be the predicted $k$-th coefficient at $T+1$. In this approach, we have
\begin{equation}\label{F-PCA}
F_{T+1}=h(\hat{Z}_{T+1})=\bar{\sigma}_{0}(m, \tau) \exp \left(\sum_{k=1}^K \hat{x}_{k}(T+1) f_{k}(m, \tau)\right).
\end{equation}

\begin{figure}[htbp!]
	\centering
	\includegraphics[scale=0.4]{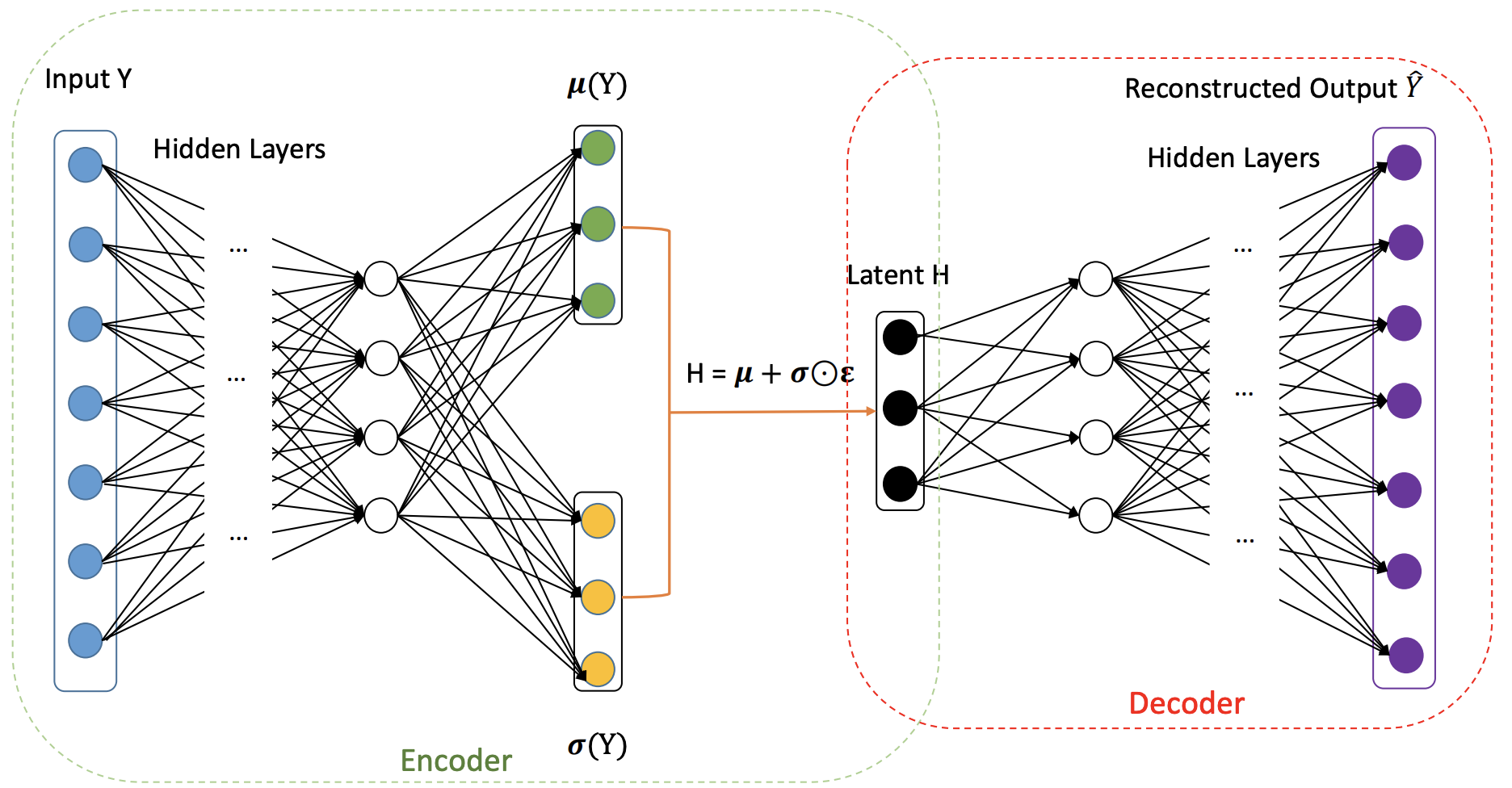}
	\caption{The structure of VAE}
	\label{fig:VAE}
\end{figure}

\smallskip	
\noindent \textbf{Method 3 (VAE)}: The variational autoencoder (VAE) is proposed in \cite{kingma2013auto}. This approach extracts latent factors to represent given data through an encoder and then tries to generate synthetic data through a decoder to resemble the given data. Specifically, the method works as follows (see Figure \ref{fig:VAE} for a graphical illustration). 
\begin{itemize}
	\item Let $Y$ be the input data vector and $H$ be the vector of $d$ latent variables. The components of $H$ are independent and $H$ follows a multivariate normal distribution with mean vector $\mu(Y)$ and standard deviation vector $\sigma(Y)$. 
	
	\item The encoder is modeled by a feedfoward neural network (FNN), denoted by $N_E$. $\mu(Y)$ and $\sigma(Y)$ are ouputs of $N_E$.  
	
	\item $H = \mu(Y) + \sigma(Y)\odot\epsilon$, where $\epsilon \sim \mathcal{N}(0, I_{d})$ with $I_d$ as the $d$-by-$d$ identity matrix and $\odot$ is the Hadamard product. 
	
	\item The decoder is modeled by another feedfoward neural network (FNN), denoted by $N_D$, with $H$ as the input. The output $\hat{Y}=N_D(H)$.	
\end{itemize}

The loss function for training the VAE has two parts. The first part is the mean squared loss between the synthetic data and the original data given by
\begin{equation}
	\mathrm{RE}=\frac{1}{M} \sum_{i=1}^{M}\left(Y_{i}-\hat{Y_{i}}\right)^{2}.
\end{equation}
where $M$ is the batch size. The second part is the Kullback-Leibler (KL) divergence between the parameterized normal distribution and $N(0,I)$ given by
\begin{equation}
		\mathrm{KL}=\frac{1}{2} \sum_{k=1}^{d}\left(-1-\log \sigma_{k}^{2}+\sigma_{k}^{2}+\mu_{k}^{2}\right)
\end{equation}
where $\mu_{k}$ and $\sigma_{k}$ are the mean and standard deviation of the $k$-th latent variable. The loss function is defined as
\begin{equation}
		\mathcal{L}(\theta_{VAE}, F) = \mathrm{RE} + \beta\mathrm{KL}.
\end{equation}
Adding the KL divergence term encourages the model to encode a distribution that is as close to normal as possible and the hyperparameter $\beta$ measures the extent of regularization. 

In our problem, $Y_t=\bar{\Sigma}_t$ and we set $Z_t=\mu(Y_t)$. With the predicted $\hat{Z}_{T+1}$, $F_{T+1}=h(\hat{Z}_{T+1})=N_D(\hat{Z}_{T+1})$, i.e., the $h$ function is given by the decoder FNN. 

\begin{remark}\label{remk:parametric-models}
A natural way to extract features from the IVS data is using a model for single day interpolation and extrapolation surveyed at the beginning of Section \ref{sec:intro}. For example, one can treat the parameters in parametric models like the surface SVI model in \cite{gatheral2014arbitrage} as features for the IVS. One advantage of using such a parametric model is that its parameters can have intuitive meanings that are easily understood by traders (\cite{bloch2019neural}). In our study, we  
calibrate the surface SVI model to our training data. However, there are two issues with predicting these calibrated parameters. First, they seem to be too volatile to be predicted well in our long training period. Second, certain constraints ensuring absence of arbitrage are not satisfied after prediction. The second one is probably a lesser issue as no arbitrage can be restored using the Step 2 DNN model in our framework. However, one cannot resolve the first issue easily. There might also be catches in using other single day models. For instance, the B-spline model of \cite{fengler2015semi} is accurate for interpolating and extrapolating the surface on a single day. One can use the control net of the B-spline model as features. However, it may vary considerably from day to day, making it difficult to predict. For the reasons above, we do not pursue these ideas for extracting features further in this paper. 
\end{remark} 

\subsection{Feature Prediction}\label{sec:feature-pred}
To predict $Z_{T+1}$ from $\{Z_0,\cdots,Z_T\}$, one can consider all kinds of models. In our experiment, we will use the long short-term memory (LSTM) model (\cite{hochreiter1997long}), which is a popular deep learning model for sequential data prediction and its success has been demonstrated in many problems. We use the model in the following way for our problem: 
\begin{itemize}
	\item For any $T$, consider
	\begin{equation}
		Z_{T}^{1}=\frac{1}{22}\sum_{t=T-21}^{T} Z_{t},\ Z_{T}^{2}=\frac{1}{5}\sum_{t=T-4}^{T}Z_{t},\ Z_{T}^{3}=Z_{T}.
	\end{equation}
	The first two represent monthly and weekly moving averages at $T$, respectively. We predict $Z_{T+1}$ using $Z^1_T, Z^2_T, Z^3_T$, which can be viewed as long-term, medium-term and short-term features. A similar approach is taken by \cite{corsi2009simple} and \cite{chen2019forecasting}.
	
	 
	\item Let $h_j$ be a hidden state that represents a summary of information from $\{Z_T^1,\cdots,Z_T^j\}$. Set $h_{0}=0$. For $j=1,2,3$, calculate
	\begin{align}
		r_{j} &=\sigma_{g}\left(W_{r} Z_{T}^j+U_{r} h_{j-1}+b_{r}\right), \\
		i_{j} &=\sigma_{g}\left(W_{i} Z_{T}^j+U_{i} h_{j-1}+b_{i}\right), \\
		o_{j} &=\sigma_{g}\left(W_{o} Z_{T}^j+U_{o} h_{j-1}+b_{o}\right), \\
		g_{j} &=\sigma_{h}\left(W_{g} Z_{T}^j+U_{g} h_{j-1}+b_{g}\right), \\
		c_{j} &=r_{j} \odot c_{j-1}+i_{j} \odot g_{j},\\
		h_{j} &=o_{j} \odot \sigma_{h}\left(c_{j}\right),\\
		y_{j} &=\sigma_{h}\left(W_{y} h_{j}+b_{y}\right).
	\end{align}
	All $W$, $U$, $b$ are parameters and $\sigma_{g}$, $\sigma_{h}$ are the sigmoid function and the tanh function, respectively, for activation. At $j$, $i_{j}$, $r_{j}$ and $o_{j}$ represent input gate, forget gate, and output gate.

	\item Finally, we predict $Z_{T+1}$ as 
	\begin{equation}
		\hat{Z}_{T+1}=\sigma_{out}(W_{out}y_3+b_{out}).
	\end{equation}	
The range of $Z_{T+1}$ varies in our framwork depending on the feature extraction method. For SAM, we use relu for $\sigma_{out}$ as $Z_{T+1}$ is positive. For VAE and PCA, since $Z_{T+1}$ can take any real value, we don't use any nonlinear activitation function and simply set $\sigma_{out}$ as the identity function.
\end{itemize}
In the following, we will write the feature prediction model as 
\begin{equation}
	\hat{Z}_{T+1}=p(Z^1_T,Z^2_T,Z^3_T;\theta_\mathcal{P}),
\end{equation}
where $\theta_\mathcal{P}$ is the vector of parameters involved. 

\subsection{The DNN Model for Surface Construction}
\begin{figure}[htbp!]
	\centering
	\includegraphics[scale=0.5]{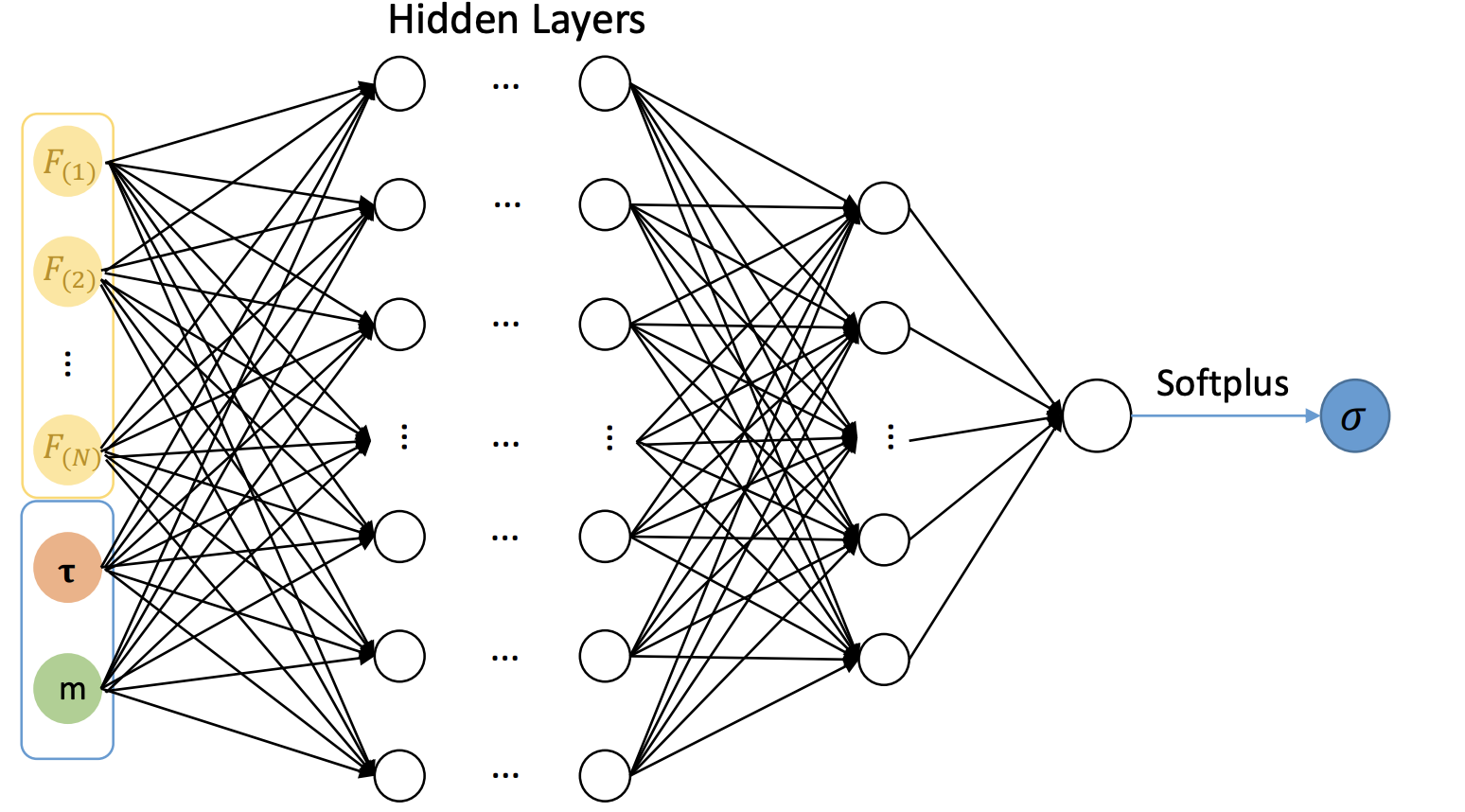}
	\caption{The DNN model for implied volatility surface construction}
	\label{fig:DNN}
\end{figure}

With $F_{T+1}=h(\hat{Z}_{T+1})$, we construct the entire implied volatility surface from $F_{T+1}$ using a DNN illustrated in Figure \ref{fig:DNN}. The neural network is a feedforward one with inputs $F_{T+1}, m, \tau$. The output is an implied volatility for the input $(m,\tau)$ pair. We use the Softplus function $\ln(1 + \exp(x))$ as the activation function of the output layer, because it makes the output nonnegative and twice differentiable, so that the first two no-arbitrage conditions in Proposition \ref{prop: NA} are fulfilled.    

\subsection{The Loss Functions and No-Arbitrage Conditions}\label{sec:loss}
Suppose the time horizon in our data is given by $\mathcal{T} = \left\{1,2,...,T\right\}$ and let $q_t$ be the number of observed implied volatilities on the surface at $t$ (in our data $q_t=374$ for all $t$ but in general it could change over time). The loss function of the featuer prediction part is given by
\begin{equation}
	\mathcal{L}_{\mathcal{P}}(\theta_{\mathcal{P}}) = \frac{1}{T} \sum_{t=1}^{T} \left\|z_{t}-p\left(Z^1_T,Z^2_T,Z^3_T; \theta_{\mathcal{P}}\right) \right\|^{2}.  \label{loss: seq}
\end{equation}
We minimize $\mathcal{L}_{\mathcal{P}}(\theta_{\mathcal{P}})$ to train the LSTM model. For the construction of the implied volatility surface, one can set the loss function as
\begin{equation}
\mathcal{L}_{\mathcal{S}}(\theta_{\mathcal{S}}) = \frac{1}{T} \sum_{t=1}^{T}\frac{1}{q_t}\sum_{i=1}^{q_t}
	(f\left(m_i, \tau_i, F_{t} ; \theta_{\mathcal{S}}\right) - \sigma_{t}\left(m_i, \tau_i \right))^{2}  \label{loss: smooth1}
\end{equation}
However, minizing $\mathcal{L}_{\mathcal{S}}(\theta_{\mathcal{S}})$ to train the surface construction model cannot guarantee the output surface is arbitrage free. By design the output of the DNN model satisfies the first two conditions in Proposition \ref{prop: NA}, but does not necessarily fulfil the other three.  Inspired by \cite{zheng2019gated} and \cite{deep_IVS}, we incorporate Conditions 3,4,5 for no arbitrage into our training by formulating them as penalties in the loss function. 

First, we create the following synthetic grids to faciliate the calculation of the penalty functions:
\begin{align}
\mathcal{I}_{\mathrm{C} 34}&=\left\{(m, \tau): m \in\left\{x^{3}: x \in\left[-\left(-2 m_{\min }\right)^{1 / 3},\left(2 m_{\max }\right)^{1 / 3}\right]_{40}\right\}, \tau \in \mathcal{T}_1\right\},\\
\mathcal{I}_{\mathrm{C} 5}&=\left\{(m, \tau): m \in\left\{6 m_{\min }, 4 m_{\min }, 4 m_{\max }, 6 m_{\max }\right\}, \tau \in \mathcal{T}_1\right\},
\end{align}
where $m_{\min}=\log(0.6)$, $m_{\max}=\log(2)$, $[a,b]_{40}$ means a uniform grid over the interval $[a,b]$ with it divided into $40$ equal parts, 
\begin{equation}
	\mathcal{T}_1=\left\{\exp (x): x \in\left[\log (1 / 365), \max \left(\log \left(\tau_{\max}+1\right)\right)\right]_{40}\right\},
\end{equation}
and $\tau_{\max}=730/365$. The grid $\mathcal{I}_{\mathrm{C} 34}$ is used for the penalty calculation associated with Condition 3 and 4 while $\mathcal{I}_{\mathrm{C} 5}$ is used for Condition 5. These grids are different from the $(m,\tau)$-grid in \eqref{sample-grid} used for sampling. In particular, $\mathcal{I}_{\mathrm{C} 34}$ has 1600 points which is a lot more than the 154 points on the sampling grid and it also covers a much wider range for both $m$ and $\tau$. We use such a dense grid on a wide region to reduce the chance of missing points on the surface at which there is significant violation of the no-arbitrage conditions. As Condition 5 considers the large moneyness behavior, we analyze moneyness levels which are extremely negative or positive.

We denote by $\mathcal{L}_{\mathrm{C}j}(\theta_{\mathcal{S}})$ the penalty function for the $j$-th condition ($j=3,4,5$). For Conditions 3 and 4, they are given by
\begin{align}
	\mathcal{L}_{\mathrm{C} 3}(\theta_{\mathcal{S}})&=\frac{1}{T}\sum_{t=1}^{T}\frac{1}{\left|\mathcal{I}_{\mathrm{C} 34}\right|}\sum_{\left(k_{i}, \tau_{i}\right) \in \mathcal{I}_{\mathrm{C} 34}} \max \left(0,-\ell_{\mathrm{cal}}\left(m_{i}, \tau_{i}, F_t;\theta_{\mathcal{S}}\right)\right), \label{cal}\\
	\mathcal{L}_{\mathrm{C} 4}(\theta_{\mathcal{S}})&=\frac{1}{T}\sum_{t=1}^{T}\frac{1}{\left|\mathcal{I}_{\mathrm{C} 34}\right|}\sum_{\left(k_{i}, \tau_{i}\right) \in \mathcal{I}_{\mathrm{C} 34}} \max \left(0,-\ell_{\mathrm{but}}\left(m_{i}, \tau_{i}, F_t;\theta_{\mathcal{S}}\right)\right),  \label{but}
\end{align}
where $\ell_{\mathrm{cal}}\left(m_{i}, \tau_{i}, F_t;\theta_{\mathcal{S}}\right)$ and $\ell_{\mathrm{but}}\left(m_{i}, \tau_{i}, F_t;\theta_{\mathcal{S}}\right)$ are defined as in \eqref{cal-NA} and \eqref{but-NA} with $\sigma$ replaced $f$. For Condition 5, it is equivalent to that the second-order derivative of $\sigma^2(m, \tau)$ goes to zero as $|m|\to\infty$, where $\partial^{2}_{mm}\sigma^2(m, \tau) = \sigma(m, \tau)\partial^{2}_{mm}\sigma(m, \tau) +  (\partial_{m}\sigma(m, \tau))^2$. Hence, the penalty is
\begin{equation} 
	\mathcal{L}_{\mathrm{C} 5}(\theta_{\mathcal{S}})=\frac{1}{T}\sum_{t=1}^{T}\frac{1}{\left|\mathcal{I}_{\mathrm{C} 5}\right|}\sum_{\left(k_{i}, \tau_{i}\right) \in \mathcal{I}_{\mathrm{C} 5}}\left|f\left(m_{i}, \tau_{i}, F_{t};\theta_{\mathcal{S}}\right)\partial^{2}_{mm}f\left(m_{i}, \tau_{i}, F_{t};\theta_{\mathcal{S}} \right) + \left(\partial_m f\left(m_{i}, \tau_{i}, F_{t};\theta_{\mathcal{S}} \right)\right)^{2}\right|.  \label{asym}
\end{equation}

Finally, we obtain our loss function for training the DNN model as
\begin{equation}
\mathcal{L}_{\mathrm{C}}(\theta_{\mathcal{S}})=\mathcal{L}_{\mathrm{S}}(\theta_{\mathcal{S}})+\lambda(\mathcal{L}_{\mathrm{C}3}(\theta_{\mathcal{S}}) + \mathcal{L}_{\mathrm{C}4}(\theta_{\mathcal{S}}) + \mathcal{L}_{\mathrm{C}5}(\theta_{\mathcal{S}}) ) \label{L_S}
\end{equation}
for some $\lambda>0$. One could use a separate penalization parameter for each penalty, but for simplicy we assume they are the same. We minimize $\mathcal{L}_{\mathrm{C}}(\theta_{\mathcal{S}})$ to train the DNN model. In our implementation, we choose $\lambda=1$, which is used in \cite{deep_IVS} in their penalized loss function. We also tried other values for $\lambda$ and found that using $\lambda=1$ results in the smallest error for the IVS on the training data and the penalties converge to zero quickly.

\begin{remark}
	To rule out static arbitrage, conditions 3 and 4 in Proposition \ref{prop: NA} must hold for every pair of $(m,\tau)$. However, in the implementation, we cannot check them at every point in the $(m,\tau)$ space, so we consider a dense grid over a wide region (see $\mathcal{I}_{\mathrm{C} 34}$). Condition 5 specifies the limiting behavior of $\sigma^2(m,\tau)$ for $|m|\to\infty$. In our implementation, we can only check this condition for very large values of $|m|$ (see $\mathcal{I}_{\mathrm{C} 5}$). It's very unlikely that the surface from our DNN model violates these constraints at points not in $\mathcal{I}_{\mathrm{C} 34}$ or $\mathcal{I}_{\mathrm{C} 5}$ (see Figure \ref{fig:SAM-DNN-curves} for the values of these penalties on the test data, which are zero if the DNN model has been trained for a sufficient number of epochs). But to be strict, one can say our DNN model yields an IVS almost free of static arbitrage.
\end{remark}

\subsection{Simulation}
Our framework can also be used to simulate the IVS over time. We can write the feature transition equation as
\begin{equation}\label{feature-tran-eq-add}
	Z_{T+1}=p(Z^1_T,Z^2_T,Z^3_T;\theta_\mathcal{P}) + \varepsilon_{T+1},
\end{equation}
or 
\begin{equation}\label{feature-tran-eq-mul}
	Z_{T+1}=p(Z^1_T,Z^2_T,Z^3_T;\theta_\mathcal{P})\exp\left(\varepsilon_{T+1}\right) ,
\end{equation}
where $\varepsilon_{T+1}$ is the error vector at $T+1$. In 
\eqref{feature-tran-eq-add} we assume additive error and in \eqref{feature-tran-eq-mul} we assume multiplicative error. The multiplicative formulation is more convenient to use than the additive one when the positivity of $Z_{T+1}$ is required.

We assume the error process $\varepsilon_1, \varepsilon_2,\cdots$ is an i.i.d. white noise with mean zero and covariance matrix $\Sigma_{\varepsilon}$. After obtaining the estimate of $\theta_\mathcal{P}$ by minimizing the loss function $\mathcal{L}_{\mathcal{P}}(\theta_{\mathcal{P}})$, one can calculate the error vector on each day and hence obtain a sample for the errors. We can assume the error vector follows a multivariate parametric distribution $F_{\eta}$ with parameter vector $\eta$ (e.g., Gaussian) and estimate $\eta$ from the error sample. 

The simulation of $Z_{T+1}$ given the available information at $T$ consists of the following steps. 
\begin{itemize}
	\item Step 1: Calculate $p(Z^1_T,Z^2_T,Z^3_T;\theta_\mathcal{P})$. 
	
	\item Step 2: Simulate $\varepsilon_{T+1}$ from $F_{\eta}$ or by bootstrapping from the error sample. 
	
	\item Step 3: Calculate $Z_{T+1}$ by \eqref{feature-tran-eq-add} or \eqref{feature-tran-eq-mul}.
	
	\item Step 4: Calculate $\sigma_{T+1}(m,\tau)=f(m,\tau,h(Z_{T+1}))$.
\end{itemize}
The DNN model $f$ ensures that the output IVS is free of static arbitrage.

\section{Empirical Results}\label{sec:empirical}
Recall that our dataset consists of daily implied volatilities for  S$\&$P 500 index options from January 1, 2009 to December 31, 2020 with a total of 3021 trading days. We split the data into training and test sets. The training dataset is from January 1, 2009 to June 27, 2018 (about $9.5$ years) while the test dataset is from June 28, 2018 to December 31, 2020 (about $2.5$ years). In particular, the US stock market crash in 2020 due to the COVID-19 pandemic is included in the test period. On each day, we observe implied volatilities for 374 pairs of $(m,\tau)$. As we only have a limited amount of training data (about 2390 days), we do not further partition it to create a validation set for hyperparameter tuning.

\subsection{Feature Extraction}
The details of the three feature extraction methods can be found in Section \ref{sec:feature}. For each day in the dateset, we extract features using these three methods and some details are as follows:
\begin{itemize}
	\item For SAM, we use $\bar{\Sigma}_t$ as the feature vector (see \eqref{sample-IVS}), which is a set of implied volatilities on a $(m,\tau)$-grid with 154 points, to represent the entire surface at $t$.   
	
	\item For PCA, we follow \cite{cont2002dynamics} to use the first three eigenmodes (i.e., $K=3$ in \eqref{F-PCA}), which already explain over $96\%$ of the variations in our data.  
	
	\item For VAE, the FNNs for the encoder and the decoder both have three hidden layers with 128 nodes per layer. We try five values for the latent dimension $d$: $2, 5, 10, 15, 20$. Their performance on the test data is shown in Table \ref{tab:VAE-d} and the difference is small, indicating the performance of the VAE model is quite robust to the choice of $d$. The VAE model with $d = 10$ achieves the smallest out-of-sample prediction error.  
\end{itemize}

\subsection{Training of the LSTM and DNN Models}
We use the LSTM model to predict the extracted features as discussed in Section \ref{sec:feature-pred}. For the DNN model for surface construction, we use three hidden layers with fifty neurons on each layer. In the training of both models, we do the following:
\begin{itemize}
	\item We initialize the parameters using Xavier initialization (\cite{glorot2010understanding}), which can prevent initial weights in a deep network from being either too large or too small. This method sets the weight of the $j$-th layer to follow a uniform distribution given by 
	\begin{equation}	
	W^{j} \sim U\left[-\frac{1}{\sqrt{n_{j}}}, \frac{1}{\sqrt{n_{j}}}\right], 
	\end{equation}
	where $n_j$ is the number of neurons on the $j$-th layer. 
	
	\item We use the Adam optimizer with minibatches (\cite{kingma2014adam}) to minimize the loss function. Calculating the gradient of the loss function using all the samples can be computationally expensive, so in each iteration we only use a minibatch (i.e., subset) of samples for the gradient evaluation. The Adam optimizer is a popular gradient-descent algorithm, which utilizes the exponentially weighted average of the gradients to accelerate convergence to the minimum.

	\item We apply batch normalization to the inputs of the neural network (\cite{ioffe2015batch}). For all the samples in a minibatch, we first estimate the mean and standard deviation of each input in this minibatch and then normalize it by subtracting its estimated mean and dividing by its estimated standard deviation. 
	
	\end{itemize}

Values of the hyperparameters associated with training and the hidden size of the models (i.e., number of neurons on a hidden layer) are displayed in Table \ref{tab:hyperparameter}. 
We train LSTM and DNN for 200 and 20 epochs, respectively. An epoch consists of all the iterations required to work through all the samples in the training set, so it is given by the size of the training data divided by the size of a minibatch.
\begin{table}[htbp!]
	\centering
	\begin{tabular}{cccccc}
		\hline
		& Epochs  &  Batch size    & Hidden size & Learning rate \\
		\hline
		LSTM&   200   & 128      & 12 & 0.01\\
		DNN& 20    &1024    & 50 & 0.001 \\
		\hline
	\end{tabular}%
	\caption{Hyperparameters for the LSTM and DNN model}
	\label{tab:hyperparameter}
\end{table}

\begin{figure}[htbp!]
	\centering
	\includegraphics[scale=0.6]{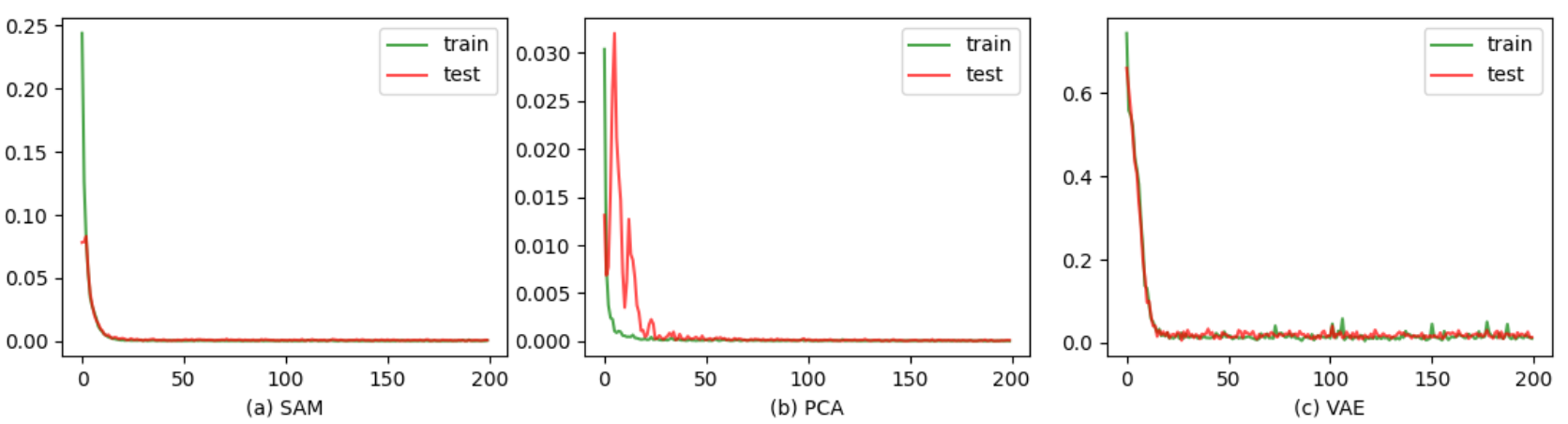}
	\caption{The loss function of the LSTM model. The $x$ variable in each plot is the epoch index.}
	\label{fig:LSTM-curves}
\end{figure}

\begin{figure}[htbp!]
    \centering
    \includegraphics[scale=0.6]{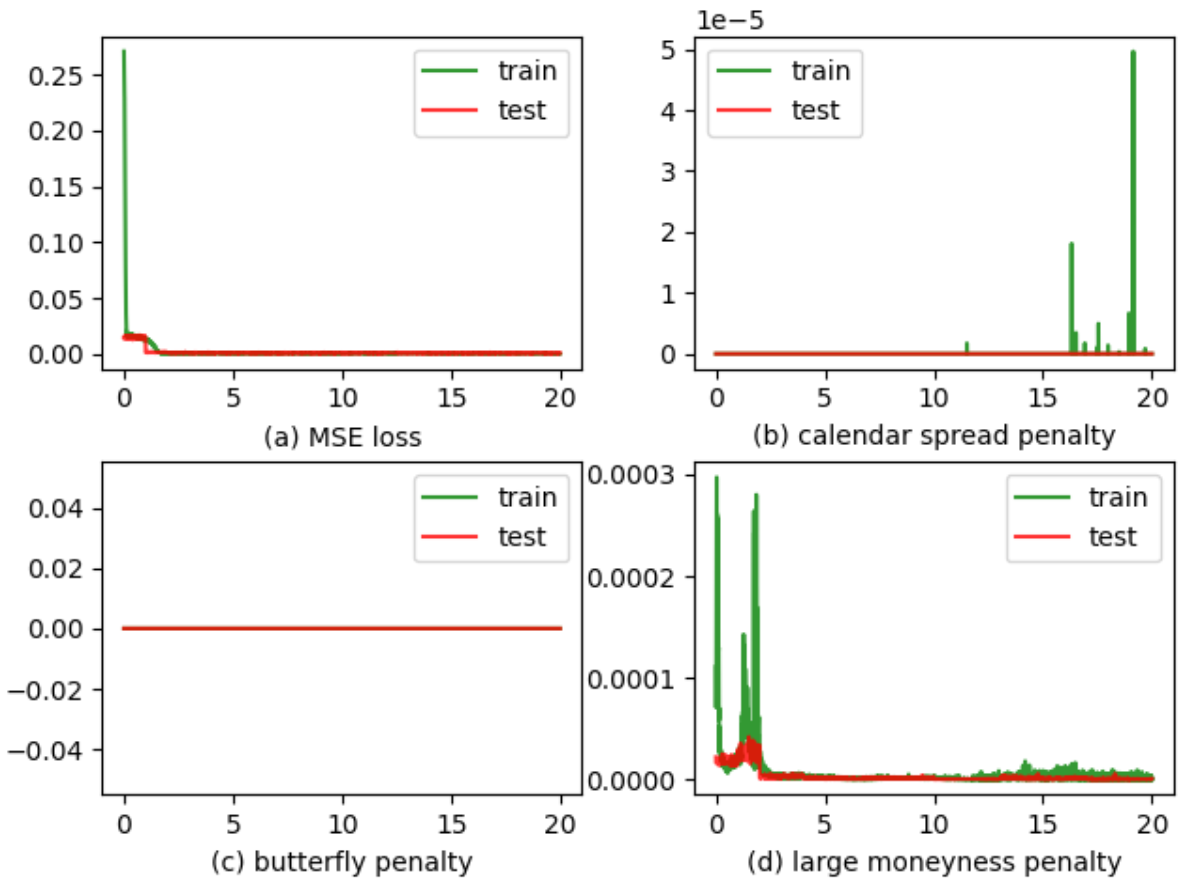}
    \caption{The MSE loss and penalties for three no-arbitrage conditions for the DNN model with features extracted from the sampling approach. The $x$ variable in each plot is the epoch index.}
    \label{fig:SAM-DNN-curves}
\end{figure}

Figures \ref{fig:LSTM-curves} and \ref{fig:SAM-DNN-curves} show the results of loss on the training data and test data as the number of epochs increases. In Figure \ref{fig:SAM-DNN-curves}, we only plot the DNN results for the model with features extracted from the sampling approach and results for the other two feature extraction approaches are similar. From Figure \ref{fig:LSTM-curves}, we can see that there is no overfitting for the LSTM model as the test loss is close to the training loss. Similarly, there is no overfitting for the DNN model as shown by Figure \ref{fig:SAM-DNN-curves}(a). The value of the penalty for three no-arbitrage conditions also become zero in the test data eventually, so there is no violation of these conditions on the synthetic grids. It should be noted that although there are some spikes for the calender spread penalty in the training data, the largest value is still very small, so the violation is insignificant.

\subsection{Out-of-Sample Prediction and Model Comparison}
Let $\hat{\theta}_{\mathcal{P}}$ and $\hat{\theta}_{\mathcal{S}}$ be the estimated parameters from the training data for the LSTM model and the DNN model, respectively.  Suppose the time index of the last day in the training period is $T_{\text{train}}$ and of the last day in the whole dataset is $T_{\text{total}}$. Set $T_{\text{test}}=T_{\text{total}}-T_{\text{train}}$, which is the number of days in the test period. We do out-of-sample test as follows: for every $t>T_{\text{train}}$, 
\begin{itemize}
	\item obtain $\hat{Z}_{t}=p(Z^1_{t-1},Z^2_{t-1},Z^3_{t-1};\hat{\theta}_\mathcal{P})$ and $F_t=h(\hat{Z}_t)$;
	\item calculate $\hat{\sigma}_{t}(m,\tau)=f(m,\tau,F_{t})$ for $(m,\tau)\in\mathcal{I}_{d,t}$. 
\end{itemize}
Here, $\mathcal{I}_{d,t}$ is the set of $(m,\tau)$-pairs in the observed implied volatility data at $t$, which contains 374 points. It is important to note that it is different from $\mathcal{I}_0$, the set of $(m,\tau)$-pairs used for sampling the surface, which has only 154 points. The error for a pair of $(m,\tau)$ is given by
\begin{equation}
	\hat{\sigma}_{t}(m,\tau)-\sigma_{t}(m,\tau)
\end{equation}
where $\sigma_{t}(m,\tau)$ is the observed implied volatility at $t$ for this pair (the ground truth). The error not only reflects the prediction error of the LSTM model for the features, but also the interpolation error of the DNN model.

To evaluate the overall out-of-sample prediction performance, we consider two commonly used error measures: root mean squared error (RMSE) and the mean absolute percentage error (MAPE). They are defined as
\begin{align}
\text{RMSE}&=\sqrt{\frac{1}{\sum_{t=1}^{T_{\text{test}}}|\mathcal{I}_{d,t}|} \sum_{t=1}^{T_{\text{test}}} \sum_{(m,\tau)\in\mathcal{I}_{d,t}}\left(\sigma_{t}(m,\tau)-\hat{\sigma}_{t}(m,\tau)\right)^{2}},\\
\text{MAPE}&=\frac{1}{\sum_{t=1}^{T_{\text{test}}}|\mathcal{I}_{d,t}|} \sum_{t=1}^{T_{\text{test}}} \sum_{(m,\tau)\in\mathcal{I}_{d,t}}\left|\frac{\sigma_{t}(m,\tau)-\hat{\sigma}_{t}(m,\tau)}{\sigma_{t}(m,\tau)}\right|.
\end{align}
In our data, $\mathcal{I}_{d,t}$ contains different $(m,\tau)$ pairs for a different $t$, but $|\mathcal{I}_{d,t}|=374$ for all $t$. \footnote{One should be cautious in comparing the errors reported in different papers. Some papers only evaluate the error on a limited set of $(m,\tau)$ pairs. For example, \cite{chen2019forecasting} only consider the errors at 45 points in the $(m,\tau)$ space.}

We examine various models. In the first step, there are three feature extraction approaches: SAM, PCA and VAE. In the second step, we consider two methods: the DNN model and the DFW model in \eqref{poly} applied to $F_t$ to predict $\sigma_t(m,\tau)$ at time $t$. The DNN model yields an arbitrage free surface whereas the DFW interpolation model cannot. This leads to six models for comparison: SAM-DNN, SAM-DFW, PCA-DNN, PCA-DFW, VAE-DNN, VAE-DFW. We also consider a classical benchmark given by the DFW model. At time $T$, we simply forecast the IVS at $T+1$ from the DFW model with its coefficients given by their estimates at $T$. 


The performance of these models on the test dataset is shown in Table \ref{tab:err}. For any pair of models, we also perform the Diebold-Mariano (DM) test (\cite{diebold2002comparing}) to assess the statistical significance of the difference in the forecast performance as measured by RMSE and the p-value is shown in Table \ref{tab:DM}. Consider model 1 and model 2. In the DM test, the null hypothesis is that the forecast error is equal while the alternative hypothesis is that the forecast error of model 1 is less than model 2. Table \ref{tab:DM} should be read in the following way. For any entry of the table, the model on its row is model 1 and the model on its column is model 2. We consider 1\% as the significance level.

Several observations can be made from Table \ref{tab:err} and \ref{tab:DM}. (1) The best performers in out-of-sample prediction are SAM-DNN and VAE-DNN. The DM test shows their difference is statistically insignificant, so they can be considered as equally good. Both of them outperfom the other models with overwhelmingly strong statistical evidence. In particular, these two models constructed in the proposed two-step framework beat the classical DFW model by a large margin. (2) The out-of-sample error of SAM-DNN and VAE-DNN is only about one third of the error of PCA-DNN. This result highlights the importance of feature selection in step 1 for predicting the IVS. While the PCA approach is good for understanding the main factors that drive the IVS movements, the approximation based on a linear combination of eigensurfaces is not sufficiently accurate for predicting the IVS. In contrast, the VAE approach is more flexible as it combines the latent factors in a nonlinear way. Thus, its improvement over PCA can be expected and this is confirmed by the results. The sampling approach can be deemed as a nonparametric way to represent the surface, which can lead to a high-dimensional feature vector (in our data its dimension is 154). Thanks to the power of LSTM, we are able to predict it quite accurately. Using a powerful model like LSTM for feature prediction is key to the success of the sampling approach. (3) The DNN model for IVS construction in step 2 also makes significant contributions to improving the prediction accuracy. For each feature extraction method, using the DNN model outperforms using the DFW model for surface construction in step 2 with statistical significance. The improvement in prediction accuracy is already considerable for SAM and even more substantial for VAE. (4) It's worth noting that the SAM-DFW model also performs quite well in prediction. It's simpler than the SAM-DNN model as it uses the simple DFW model instead of the complex DNN model for surface construction in step 2.

\begin{figure}[htbp!]
	\centering
	\includegraphics[scale=0.50]{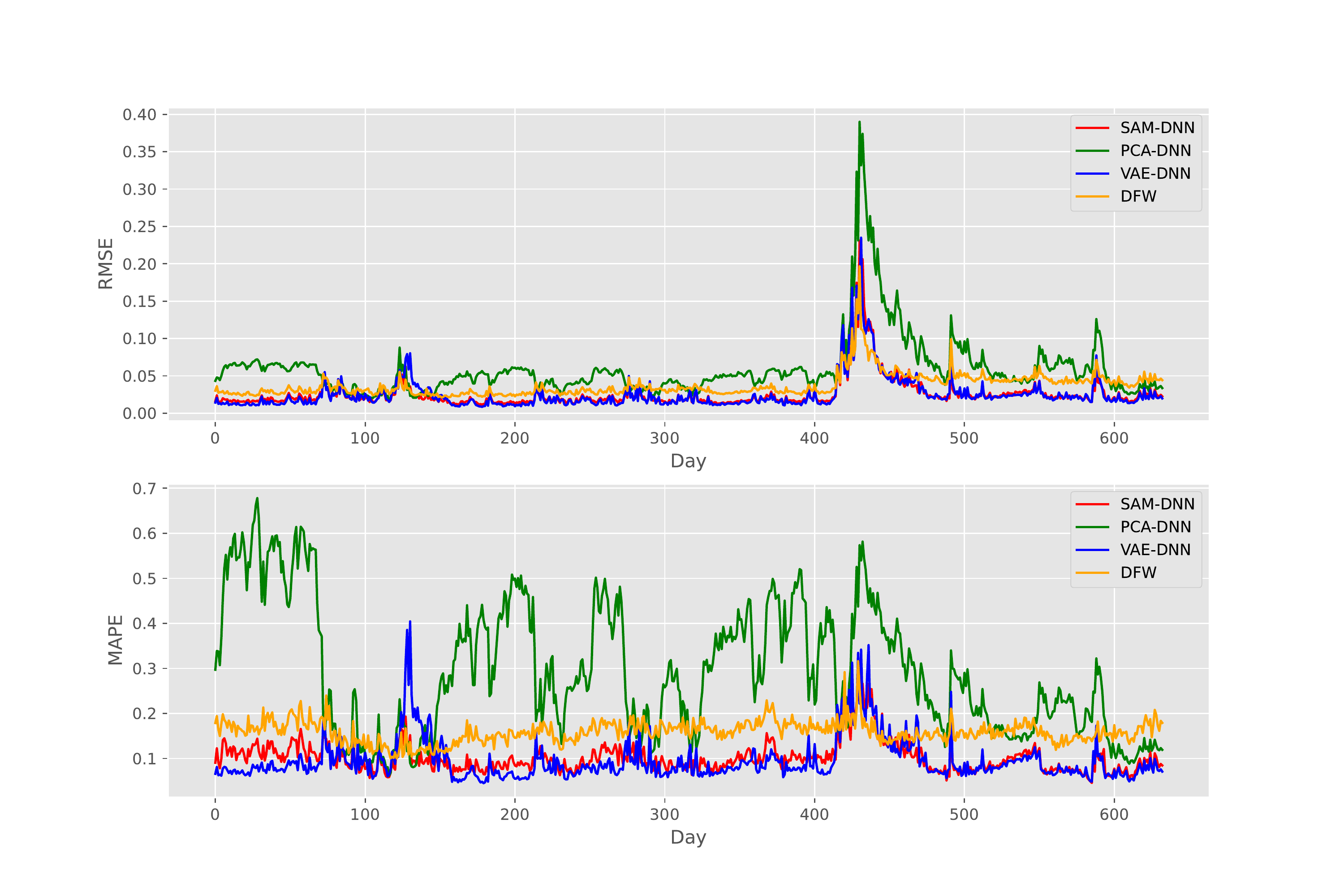}
	\caption{The daily RMSE and MAPE in the test period}
	\label{fig:err}
\end{figure}

We further plot the RMSE and MAPE of each day in the test period for four models in Figure \ref{fig:err}. Both SAM-DNN and VAE-DNN are better than PCA-DNN throughout the preiod and they also outperform DFW on most days. However, the error of all four models spikes up in March, 2020, during which the US stock market suffered a meltdown due to the pandemic. The relatively big error signals a shift in the market regime in that period. The features we use in all the models are extracted from the implied volatility data, which do not provide a direct representation of the market regime. To improve their performance, one can further augment the feature vector with exogenous variables like index return and VIX which are proxies of the market regime. 

\begin{table}[htbp!]
	\centering
	\begin{tabular}{cccccccc}
		\hline
		& SAM-DNN      & SAM-DFW &PCA-DNN          & PCA-DFW&VAE-DNN &VAE-DFW& DFW \\
		\hline
		Training set&         &          &            &             &          &        & \\
		RMSE         & $\textbf{0.0202}$& 0.0288   &0.0527        & 0.0608 &0.0205  & 0.0633 & 0.0346 \\
		MAPE         & 7.98\%  & 11.65\% &27.65\%    & 27.88\% &$\textbf{7.60\%}$ & 34.21\%  & 14.83\%\\
		
		\hline
		Test set  &        &       &         &             &             &          &     \\
		RMSE         & $\textbf{0.0245}$ & 0.0312    &0.0544    & 0.0745&0.0248      &0.0647 &   0.0366\\
		MAPE         & 9.90\% & 12.88\%    &28.93\%    &30.41\%&$\textbf{9.46\%}$     &32.75\%  & 15.83\%\\
		
		\hline
	\end{tabular}%
	\caption{The RMSE and MAPE for six models. The results of VAE-DNN and VAE-DFW are for $d=10$. The smallest value on a row is highlighted in bold. }
	\label{tab:err}%
\end{table}%

\begin{table}[htbp!]
	\centering
	\begin{tabular}{cccccc}
		\hline
		& $d=2$   & $d=5$  & $d=10$ & $d=15$ & $d=20$ \\
		\hline
		Training set&         &          &            &             &          \\
		RMSE&  0.0208    &   0.0207    &  $\textbf{0.0205}$          &  0.0210        & 0.0231         \\
		MAPE&7.90\%     &  7.71\%     &    \textbf{7.60\%}        &  8.00\%           &  9.20\%        \\
		\hline
		Test set&         &          &            &             &          \\
		RMSE&  0.0262      & 0.0258     &$ \textbf{0.0248} $          &  0.0253           &  0.0268        \\
		MAPE&  9.98\%       &  9.76\%     &  $\textbf{9.46\%} $         &   9.75\%          &  10.33\%        \\
		\hline
	\end{tabular}%
	\caption{The RMSE and MAPE of the VAE-DNN model with different latent dimensions. The smallest value on a row is highlighted in bold.}
	\label{tab:VAE-d}%
\end{table}%

\begin{table}[htbp!]
	\centering
	\begin{tabular}{cccccccc}
		\hline
		& SAM-DNN      & SAM-DFW &PCA-DNN          & PCA-DFW&VAE-DNN &VAE-DFW& DFW \\
		\hline
		SAM-DNN          & -   &  $2.7\text{e-04}$  &$2.2\text{e-16}$  & $2.2\text{e-16}$ &0.5639  & $2.2\text{e-16}$& $4.9\text{e-05}$ \\
		SAM-DFW       & 0.9997  & -  &$2.2\text{e-16}$   & $2.2\text{e-16}$ &0.9999 &2.2\text{e-16}  & 0.1495\\						
		PCA-DNN        & 0.9999 & 0.9999    &-   &$2.2\text{e-16}$  &0.9999    & $5.3\text{e-08}$ &   0.9999\\
		PCA-DFW    & 0.9999 & 0.9999     &0.9999   &- &0.9999     &0.9999  & 0.9999\\
		VAE-DNN    & 0.4361& $1.8\text{e-13}$     &$2.2\text{e-16}$  &$2.2\text{e-16}$&- &$2.2\text{e-16}$   & $3.2\text{e-07}$ \\
		VAE-DFW    & 0.9999 & 0.9999    &0.9999    &$1.2\text{e-12}$    &0.9999     &- & 0.9999\\
		DFW    & 0.9999 & 0.8505    &$2.2\text{e-16}$     &$2.2\text{e-16}$ &0.9999     &$2.2\text{e-16}$  & -\\
		\hline
	\end{tabular}%
	\caption{p-value of the Diebold-Mariano test with RMSE as the error measure}
	\label{tab:DM}%
\end{table}%

For the predicted IVS on the test days, we check for violation of the constraints on calendar spread arbitrage and butterfly arbitrage. Consider
\begin{align}
	L_{cal}^-&=\frac{1}{\sum_{t=1}^{T_{\text{test}}}|\mathcal{I}_{d,t}|} \sum_{t=1}^{T_{\text{test}}} \sum_{(m,\tau)\in\mathcal{I}_{d,t}}\min(\ell_{\mathrm{cal}}(m, \tau), 0),\\
	L_{but}^-&=\frac{1}{\sum_{t=1}^{T_{\text{test}}}|\mathcal{I}_{d,t}|} \sum_{t=1}^{T_{\text{test}}} \sum_{(m,\tau)\in\mathcal{I}_{d,t}}\min(\ell_{\mathrm{but}}(m, \tau), 0).
\end{align}
The no-arbitrage constraints require $\ell_{\mathrm{cal}}(m, \tau)$ and $\ell_{\mathrm{but}}(m, \tau)$ to be nonnegative for any $(m,\tau)$. In the above quantities, we check $\ell_{\mathrm{cal}}(m, \tau)$ and $\ell_{\mathrm{but}}(m, \tau)$ on the $(m,\tau)$ grid for the observed implied volatilities for each test day. A negative value at a pair of $(m,\tau)$ indicates violation at this point and we calculate the average of the negative values over all test days. The results are reported for four models in Table \ref{tab:violation}. While no violation is detected for the three models that use DNN in step 2, prediction under the DFW model yields implied volatility surfaces with arbitrage opportunities and calendar spread arbitrage in particular. 

\begin{table}[htbp!]
	\centering
	\begin{tabular}{ccccc}
		\hline
		& SAM-DNN       &PCA-DNN     &VAE-DNN & DFW \\
		\hline
		$L_{cal}^- $       & 0.0 &0.0   &0.0&  -0.1142\\
		$L_{but}^- $      & 0.0 &0.0     &0.0 & -0.0002 \\		
		\hline
	\end{tabular}%
	\caption{Violation of no-arbitrage conditions for calendar spread and butterfly arbitrage}
	\label{tab:violation}%
\end{table}%

\begin{remark}\label{remk:PCA-OU}
	For the PCA approach, we have also tried predicting each expansion coefficient using a separate AR(1) model (autoregressive model of order one), which is used in \cite{cont2002dynamics} for modeling the dynamics of these coefficients. One can view PCA(AR)+DFW as a simple model constructed based on classical statistical techniques without using machine learning. Table \ref{tab:LSTM-AR} shows that predicting the PCA coefficients using AR and LSTM are very close in the prediction performance for the IVS.
	\begin{table}[htbp!]
		\centering
		\begin{tabular}{ccc}
			\hline
			& PCA(LSTM)+DFW   & PCA(AR)+DFW \\
			\hline
			Training set&         &          \\
			RMSE&  0.0608  &   0.0649   \\
			MAPE&27.88\%   &  28.83\%    \\
			\hline
			Test set&         &            \\
			RMSE&  0.0745       & 0.0765     \\
			MAPE&  30.41\%       &  30.47\%    \\
			\hline
		\end{tabular}%
		\caption{Performance of the PCA+DFW model using LSTM and AR(1) for feature prediction}
		\label{tab:LSTM-AR}%
	\end{table}%
\end{remark}

\section{Conclusion}\label{sec:conclusion}
We develop a flexible two-step framework for predicting and simulating the implied volatility surface dynamically free of static arbitrage. The first step involves constructing features to represent the IVS and predicting/simulating them. The second step constructs an IVS without static arbitrage through a deep neural network model from the predicted/simulated features. Using this framework, we develop two models that are quite successful in predicting the IVS. One of them extracts features by directly sampling the IVS data on a grid and the other extracts latent factors through the encoder neural network in the variational autoencoder. Both models significantly outperform a classical parametric model. 

The prediction accuracy of our models can be further improved in the following ways. First, we can add exogenous variables such as index return and VIX to the feature vector to represent the market regime. We expect that these features would boost the prediction power of our models when the marke is under stress. Second, we can use more sophisticated deep learning models for predicting the features and they might be particularly helpful if the feature vector is high dimensional and exhibits complex behavior. In future research, we also plan to apply our models to financial applications such as hedging, risk management and options trading.

\section*{Acknowledgements}
The first two authors were supported by Hong Kong Research Grant Council General Research Fund Grant 14206020. The third author was supported by National Science Foundation of China Grant 11801423 and by Shenzhen Basic Research Program Project JCYJ20190813165407555.

\bibliographystyle{chicagoa}
\bibliography{references}

\begin{thebibliography}{}

\bibitem[\protect\citeauthoryear{Ackerer, Tagasovska, and Vatter}{Ackerer
  et~al.}{2020}]{deep_IVS}
Ackerer, D., N.~Tagasovska, and T.~Vatter (2020).
\newblock Deep smoothing of the implied volatility surface.
\newblock In H.~Larochelle, M.~Ranzato, R.~Hadsell, M.~F. Balcan, and H.~Lin
  (Eds.), {\em Advances in Neural Information Processing Systems}, Volume~33,
  pp.\  11552--11563. Curran Associates, Inc.

\bibitem[\protect\citeauthoryear{Almeida, Fan, and Tang}{Almeida
  et~al.}{2021}]{almeida2021can}
Almeida, C., J.~Fan, and F.~Tang (2021).
\newblock Can a machine correct option pricing models?
\newblock {\em Available at SSRN 3835108\/}.


\bibitem[\protect\citeauthoryear{Audrino and Colangelo}{Audrino and
  Colangelo}{2010}]{audrino2010semi}
Audrino, F. and D.~Colangelo (2010).
\newblock Semi-parametric forecasts of the implied volatility surface using
  regression trees.
\newblock {\em Statistics and Computing\/}~{\em 20\/}(4), 421--434.


\bibitem[\protect\citeauthoryear{Bergeron, Fung, Poulos, Hull, and
  Veneris}{Bergeron et~al.}{2021}]{bergeron2021variational}
Bergeron, M., N.~Fung, Z.~Poulos, J.~C. Hull, and A.~Veneris (2021).
\newblock Variational autoencoders: A hands-off approach to volatility.
\newblock Available from https://dx.doi.org/10.2139/ssrn.3827447.


\bibitem[\protect\citeauthoryear{Bernales and Guidolin}{Bernales and
  Guidolin}{2014}]{bernales2014can}
Bernales, A. and M.~Guidolin (2014).
\newblock Can we forecast the implied volatility surface dynamics of equity
  options? predictability and economic value tests.
\newblock {\em Journal of Banking \& Finance\/}~{\em 46}, 326--342.


\bibitem[\protect\citeauthoryear{Bloch}{Bloch}{2019}]{bloch2019neural}
Bloch, D.~A. (2019).
\newblock Neural networks based dynamic implied volatility surface.
\newblock Available from
  https://papers.ssrn.com/sol3/papers.cfm?abstract\textunderscore id=3492662.


\bibitem[\protect\citeauthoryear{Bloch and B{\"o}{\"o}k}{Bloch and
  B{\"o}{\"o}k}{2020}]{bloch2020predicting}
Bloch, D.~A. and A.~B{\"o}{\"o}k (2020).
\newblock Predicting future implied volatility surface using {TDBP}-learning.
\newblock Available from https: //dx.doi.org/10.2139/ssrn.3739514.


\bibitem[\protect\citeauthoryear{Borovykh, Bohte, and Oosterlee}{Borovykh
  et~al.}{2017}]{borovykh2017conditional}
Borovykh, A., S.~Bohte, and C.~W. Oosterlee (2017).
\newblock Conditional time series forecasting with convolutional neural
  networks.
\newblock Avaliable from https://arxiv.org/pdf/1703.04691v1.pdf.


\bibitem[\protect\citeauthoryear{Cao, Chen, and Hull}{Cao
  et~al.}{2020}]{cao2020neural}
Cao, J., J.~Chen, and J.~Hull (2020).
\newblock A neural network approach to understanding implied volatility
  movements.
\newblock {\em Quantitative Finance\/}~{\em 20\/}(9), 1405--1413.


\bibitem[\protect\citeauthoryear{Chen and Zhang}{Chen and
  Zhang}{2019}]{chen2019forecasting}
Chen, S. and Z.~Zhang (2019).
\newblock Forecasting implied volatility smile surface via deep learning and
  attention mechanism.
\newblock Available from https://dx.doi.org/10.2139/ssrn.3508585.


\bibitem[\protect\citeauthoryear{Cont and Da~Fonseca}{Cont and
  Da~Fonseca}{2002}]{cont2002dynamics}
Cont, R. and J.~Da~Fonseca (2002).
\newblock Dynamics of implied volatility surfaces.
\newblock {\em Quantitative Finance\/}~{\em 2}, 45--60.


\bibitem[\protect\citeauthoryear{Corsi}{Corsi}{2009}]{corsi2009simple}
Corsi, F. (2009).
\newblock A simple approximate long-memory model of realized volatility.
\newblock {\em Journal of Financial Econometrics\/}~{\em 7\/}(2), 174--196.


\bibitem[\protect\citeauthoryear{Dellaportas and Mijatovi{\'c}}{Dellaportas and
  Mijatovi{\'c}}{2014}]{dellaportas2014arbitrage}
Dellaportas, P. and A.~Mijatovi{\'c} (2014).
\newblock Arbitrage-free prediction of the implied volatility smile.
\newblock Available from https://arxiv.org/abs/1407.5528.


\bibitem[\protect\citeauthoryear{Diebold and Mariano}{Diebold and
  Mariano}{2002}]{diebold2002comparing}
Diebold, F.~X. and R.~S. Mariano (2002).
\newblock Comparing predictive accuracy.
\newblock {\em Journal of Business \& Economic Statistics\/}~{\em 20\/}(1),
  134--144.


\bibitem[\protect\citeauthoryear{Dumas, Fleming, and Whaley}{Dumas
  et~al.}{1998}]{dumas1998implied}
Dumas, B., J.~Fleming, and R.~E. Whaley (1998).
\newblock Implied volatility functions: Empirical tests.
\newblock {\em The Journal of Finance\/}~{\em 53\/}(6), 2059--2106.


\bibitem[\protect\citeauthoryear{Fengler}{Fengler}{2009}]{fengler2009arbitrage}
Fengler, M.~R. (2009).
\newblock Arbitrage-free smoothing of the implied volatility surface.
\newblock {\em Quantitative Finance\/}~{\em 9\/}(4), 417--428.


\bibitem[\protect\citeauthoryear{Fengler, H{\"a}rdle, and Mammen}{Fengler
  et~al.}{2007}]{fengler2007semiparametric}
Fengler, M.~R., W.~K. H{\"a}rdle, and E.~Mammen (2007).
\newblock A semiparametric factor model for implied volatility surface
  dynamics.
\newblock {\em Journal of Financial Econometrics\/}~{\em 5\/}(2), 189--218.


\bibitem[\protect\citeauthoryear{Fengler, H{\"a}rdle, and Villa}{Fengler
  et~al.}{2003}]{fengler2003dynamics}
Fengler, M.~R., W.~K. H{\"a}rdle, and C.~Villa (2003).
\newblock The dynamics of implied volatilities: A common principal components
  approach.
\newblock {\em Review of Derivatives Research\/}~{\em 6\/}(3), 179--202.


\bibitem[\protect\citeauthoryear{Fengler and Hin}{Fengler and
  Hin}{2015}]{fengler2015semi}
Fengler, M.~R. and L.-Y. Hin (2015).
\newblock Semi-nonparametric estimation of the call-option price surface under
  strike and time-to-expiry no-arbitrage constraints.
\newblock {\em Journal of Econometrics\/}~{\em 184\/}(2), 242--261.


\bibitem[\protect\citeauthoryear{Gatheral}{Gatheral}{2004}]{gatheral2004parsimonious}
Gatheral, J. (2004).
\newblock A parsimonious arbitrage-free implied volatility parameterization
  with application to the valuation of volatility derivatives.
\newblock Presentation at Global Derivatives \& Risk Management, Madrid.


\bibitem[\protect\citeauthoryear{Gatheral and Jacquier}{Gatheral and
  Jacquier}{2014}]{gatheral2014arbitrage}
Gatheral, J. and A.~Jacquier (2014.).
\newblock Arbitrage-free {SVI} volatility surfaces.
\newblock {\em Quantitative Finance\/}~{\em 14\/}(1), 59--71.


\bibitem[\protect\citeauthoryear{Glorot and Bengio}{Glorot and
  Bengio}{2010}]{glorot2010understanding}
Glorot, X. and Y.~Bengio (2010).
\newblock Understanding the difficulty of training deep feedforward neural
  networks.
\newblock In {\em Proceedings of the thirteenth international conference on
  artificial intelligence and statistics}, pp.\  249--256. JMLR Workshop and
  Conference Proceedings.

\bibitem[\protect\citeauthoryear{Goncalves and Guidolin}{Goncalves and
  Guidolin}{2006}]{goncalves2006predictable}
Goncalves, S. and M.~Guidolin (2006).
\newblock Predictable dynamics in the s\&p 500 index options implied volatility
  surface.
\newblock {\em The Journal of Business\/}~{\em 79\/}(3), 1591--1635.


\bibitem[\protect\citeauthoryear{Gulisashvili}{Gulisashvili}{2012}]{gulisashvili2012analytically}
Gulisashvili, A. (2012).
\newblock {\em Analytically Tractable Stochastic Stock Price Models}.
\newblock Springer Science \& Business Media.


\bibitem[\protect\citeauthoryear{H{\"a}rdle}{H{\"a}rdle}{1990}]{hardle1990applied}
H{\"a}rdle, W. (1990).
\newblock {\em Applied Nonparametric Regression}.
\newblock Cambridge University Press.


\bibitem[\protect\citeauthoryear{Hochreiter and Schmidhuber}{Hochreiter and
  Schmidhuber}{1997}]{hochreiter1997long}
Hochreiter, S. and J.~Schmidhuber (1997).
\newblock Long short-term memory.
\newblock {\em Neural Computation\/}~{\em 9\/}(8), 1735--1780.


\bibitem[\protect\citeauthoryear{Horvath, Muguruza, and Tomas}{Horvath
  et~al.}{2021}]{horvath2021deep}
Horvath, B., A.~Muguruza, and M.~Tomas (2021).
\newblock Deep learning volatility: a deep neural network perspective on
  pricing and calibration in (rough) volatility models.
\newblock {\em Quantitative Finance\/}~{\em 21\/}(1), 11--27.


\bibitem[\protect\citeauthoryear{Ioffe and Szegedy}{Ioffe and
  Szegedy}{2015}]{ioffe2015batch}
Ioffe, S. and C.~Szegedy (2015).
\newblock Batch normalization: Accelerating deep network training by reducing
  internal covariate shift.
\newblock In {\em International conference on machine learning}, pp.\
  448--456. PMLR.

\bibitem[\protect\citeauthoryear{Kingma and Ba}{Kingma and
  Ba}{2014}]{kingma2014adam}
Kingma, D.~P. and J.~Ba (2014).
\newblock Adam: A method for stochastic optimization.
\newblock {\em arXiv preprint arXiv:1412.6980,
  https://doi.org/10.1016/j.asoc.2020.106181\/}.


\bibitem[\protect\citeauthoryear{Kingma and Welling}{Kingma and
  Welling}{2013}]{kingma2013auto}
Kingma, D.~P. and M.~Welling (2013).
\newblock Auto-encoding variational {B}ayes.
\newblock Available from https://arxiv.org/abs/1312.6114.


\bibitem[\protect\citeauthoryear{Ning, Jaimungal, Zhang, and Bergeron}{Ning
  et~al.}{2021}]{ning2021arbitrage}
Ning, B., S.~Jaimungal, X.~Zhang, and M.~Bergeron (2021).
\newblock Arbitrage-free implied volatility surface generation with variational
  autoencoders.
\newblock {\em arXiv preprint arXiv:2108.04941\/}.


\bibitem[\protect\citeauthoryear{Orosi}{Orosi}{2015}]{orosi2015arbitrage}
Orosi, G. (2015).
\newblock Arbitrage-free call option surface construction using regression
  splines.
\newblock {\em Applied Stochastic Models in Business and Industry\/}~{\em
  31\/}(4), 515--527.


\bibitem[\protect\citeauthoryear{Roper}{Roper}{2010}]{roper2010arbitrage}
Roper, M. (2010).
\newblock Arbitrage free implied volatility surfaces.
\newblock Available from
  https://talus.maths.usyd.edu.au/u/pubs/publist/preprints/2010/roper-9.pdf.


\bibitem[\protect\citeauthoryear{Sezer, Gudelek, and Ozbayoglu}{Sezer
  et~al.}{2020}]{sezer2020financial}
Sezer, O.~B., M.~U. Gudelek, and A.~M. Ozbayoglu (2020).
\newblock Financial time series forecasting with deep learning: A systematic
  literature review: 2005--2019.
\newblock {\em Applied Soft Computing\/}~{\em 90}, 106181.


\bibitem[\protect\citeauthoryear{Sirignano and Cont}{Sirignano and
  Cont}{2019}]{sirignano2019universal}
Sirignano, J. and R.~Cont (2019).
\newblock Universal features of price formation in financial markets:
  perspectives from deep learning.
\newblock {\em Quantitative Finance\/}~{\em 19\/}(9), 1449--1459.


\bibitem[\protect\citeauthoryear{Sirignano, Sadhwani, and Giesecke}{Sirignano
  et~al.}{2018}]{sirignano2018deep}
Sirignano, J., A.~Sadhwani, and K.~Giesecke (2018).
\newblock Deep learning for mortgage risk.
\newblock Available from https://dx.doi.org/10.2139/ssrn.2799443.


\bibitem[\protect\citeauthoryear{Sirignano}{Sirignano}{2019}]{sirignano2019deep}
Sirignano, J.~A. (2019).
\newblock Deep learning for limit order books.
\newblock {\em Quantitative Finance\/}~{\em 19\/}(4), 549--570.


\bibitem[\protect\citeauthoryear{Skiadopoulos, Hodges, and
  Clewlow}{Skiadopoulos et~al.}{2000}]{skiadopoulos2000dynamics}
Skiadopoulos, G., S.~Hodges, and L.~Clewlow (2000).
\newblock The dynamics of the {S\&P} 500 implied volatility surface.
\newblock {\em Review of Derivatives Research\/}~{\em 3\/}(3), 263--282.


\bibitem[\protect\citeauthoryear{Yan, Zhang, Ma, Liu, and Wu}{Yan
  et~al.}{2018}]{yan2018}
Yan, X., W.~Zhang, L.~Ma, W.~Liu, and Q.~Wu (2018).
\newblock Parsimonious quantile regression of financial asset tail dynamics via
  sequential learning.
\newblock In S.~Bengio, H.~Wallach, H.~Larochelle, K.~Grauman, N.~Cesa-Bianchi,
  and R.~Garnett (Eds.), {\em Advances in Neural Information Processing
  Systems}, Volume~31. Curran Associates, Inc.

\bibitem[\protect\citeauthoryear{Zeng and Klabjan}{Zeng and
  Klabjan}{2019}]{zeng2019online}
Zeng, Y. and D.~Klabjan (2019).
\newblock Online adaptive machine learning based algorithm for implied
  volatility surface modeling.
\newblock {\em Knowledge-Based Systems\/}~{\em 163}, 376--391.


\bibitem[\protect\citeauthoryear{Zheng, Yang, and Chen}{Zheng
  et~al.}{2019}]{zheng2019gated}
Zheng, Y., Y.~Yang, and B.~Chen (2019.).
\newblock Gated deep neural networks for implied volatility surfaces.
\newblock Available from https://arxiv.org/pdf/1904.12834.pdf.


\end{thebibliography}

\end{document}